# Citation Recommendation based on Argumentative Zoning of User Queries

Shutian Ma[1,2], Chengzhi Zhang[1,*], Heng Zhang[1], Zheng Gao[3]

1. Department of Information Management, Nanjing University of Science and Technology, Nanjing, 210094, China

2. Tencent, Shenzhen, 518063, China

3. Department of Information & Library Science, Indiana University Bloomington, 47405, USA

**Abstract**：Citation recommendation aims to locate the important papers for scholars to cite. When writing the citing sentences, the authors usually hold different citing intents, which are referred to citation function in citation analysis. Since argumentative zoning is to identify the argumentative and rhetorical structure in scientific literature, we want to use this information to improve the citation recommendation task. In this paper, a multi-task learning model is built for citation recommendation and argumentative zoning classification. We also generated an annotated corpus of the data from PubMed Central based on a new argumentative zoning schema. The experimental results show that, by considering the argumentative information in the citing sentence, citation recommendation model will get better performance.

## 1. Introduction

Due to the increasing of scientific publication, scientific information recommendation has become an urgent problem which can save retrieval cost. There are kinds of information that can be recommended, such as paper recommendation (Mei et al., 2022) , author recommendation (Alhoori & Furuta, 2017), journal recommendation (Gündoğan et al., 2023) and so on. Among them, citation recommendation has arisen researchers' attention, which aims to help people find appropriate and necessary work to cite based on the given user queries. This paper aims to improve citation recommendation by considering the argumentative zoning of the citing sentence. Normally，authors will follow a logical framework when writing scientific papers. For example, the International Committee of Medical Journal Editors (ICMJE) recommends the IMRaD (Introduction, Methods, Results and Discussion) structure in writing and editing guidelines of biomedical publications (Editors & others, 2004). The structure of a research article is designed to present the research work clearly and concisely. This structure also helps to make it easy for readers to understand and evaluate the research. Previous studies have classified the scientific texts based on argumentative zoning (Hua et al., 2019; Liakata et al., 2010). Except using a certain sequence of arguments, authors need to state their point of view clearly. When people cite papers, they will have various aims. For example, Figure 1. shows two citing sentences with different purposes in the paper titled with Context-aware Citation Recommendation (He et al., 2010). The first citing sentence is intended to describe the results that observed in previous study. The second citing sentence wants to introduce the research goal and content. Obviously, there are different rhetorical structures or citation functions between these two sentences.

| Citing Paper |
| --- |
| Title: Context-aware Citation Recommendation |
| For example, Huang et al. [CITE] observed that citation contexts can effectively help to avoid "topic drifting" in clustering citations into topics. |
| Ritchie [CITE] extensively examined the impact of various citation Context extraction methods on the performance of information Retrieval. |

The first citation is intended to give the *results* that observed in previous study (Huang et al.)

The second citation is intended to give an *introduction* of the research (Ritchie).

**Figure 1. Example of citations with different aims**

To figure out if such semantic information would improve the citation recommenders or not, this paper conducts argumentative zoning during citation recommendation. We firstly set classification schema according to the previous work related with argumentative zoning and citation function. Then, we conducted annotations of the citing sentence and its corresponding citation papers in PubMed dataset. After the agreement estimation over labeled data, our recommendation model was trained and did prediction over testing data. Experiments showed that the performance of citation recommendation would get better when considering the argumentative information of citing sentences.

This paper is then organized as follows. Section 2 provides a brief review of citation recommendation, argumentative zoning, and citation intents. Section 3 introduces the detailed information about our corpus construction and proposed model. Experiment and results are illustrated in Section 4. Conclusion and future work are given in Section 5.

## 2. Related Work

In this section, we will introduce related work from three topics: citation recommendation, argumentative zoning of scientific texts and citation function classification.

### 2.1. Citation Recommendation

Given the user query and set of published papers, citation recommendation is a task of recommending suitable citations to scholars based on queries, which can help them improve searching efficiently (S. Ma et al., 2020). Basically, there are two kinds of citation recommendations (He et al., 2010). One is global citation recommendation which is to recommend a list of references based on the manuscript provided by users. Another one is local citation recommendation which is to recommend papers according to the specific citation context. Such citation context usually contains several sentences around a placeholder, like ''[]''

(Färber & Jatowt, 2020). Meanwhile, except the explicit citations (in the form of author name and paper year, or using a bracketed notation), researchers also investigate the detection of implicit citations (AbuRa'ed et al., 2018). In this paper, we will focus on the explicit citation and solve the problem of local citation recommendation.

At first, citation recommendations concentrate on how to narrow the semantic difference between the citation context and the candidate papers (Ali et al., 2021). Tang and Zhang (2009) proposed the RBM-CS model to learn a mixture of topic distribution over the relationships between citation and paper contents. In 2010, a non-parametric probabilistic model was designed for measuring the context-based relevance between the citation context and a document in the CiteSeerX system (He et al., 2010). With the enrichment of meta data and development of graph algorithms, relationship between papers, authors and venues are then applied. Dai et al. (2018) considered not only text content similarity between papers but also community relevance among authors for effective recommendation. Yang et al. (2019) constructed a heterogeneous bibliographic network which contained nodes such as papers and authors.

Lately, neural network attracted lots of attention in building citation recommenders. Yang et al. (2019) developed an attention-based encoder-decoder (AED) model for local citation recommendation. Wang et al. (2020) constructed Bi-LSTM model to learn the representations of papers and citation contexts, and the author information and citation relationship were integrated in the vector representations. Jeong et al. (2020) proposed a model which comprises a document encoder and a context encoder to learn the information of textual data and graph data contained in papers. Pornprasit et al. (2022) developed ConvCN which learns the citation network representations to enhance graph-based citation recommendation algorithms. In order to find the most appropriate form of citation content, Zhang and Zhu (2022) designed four kinds citation content: current sentences, which is the most direct text that related to the citation motivation; current sentences and surrounding sentences (CS&SS); current sentences and surrounding sentences that do not cite other references; automatic summarization of CS&SS. H.-C. Wang et al. (2022) developed SentCite to identify the citing sentences and conduct citation recommendation. Muther and Smith (2023) fine-tuned a variety of language models to rerank candidate content sources. Yin, Wang, and Ling (2024) trained a pair of neural network encoders that map citation contexts and all possible cited papers to the same vector space to do citation recommendation. Liu, Chen, Lee, and Huang (2024) utilized the academic networks to generate the evolving knowledge graph embedding which helps improving the recommendation model accuracy.

As we can see, researchers are paying attention to textual representation over scientific contents and node embedding over citation networks. By building different structures of deep learning models, content-based or graph-based information are integrated in neural network. Actually, more sematic information could be considered, for example, the rhetorical structures hidden behind citing sentences that we mentioned in introduction part.

## 2.2. Argumentative Zoning of Scientific Text

Recently, there is a literature survey about argument mining task (Al Khatib et al., 2021). As it is summarized, relevant research can be categorized under four areas of study, which are corpus creation and new annotation schemes (Lauscher et al., 2018; Toulmin, 2003; Walton et al.,

2008), automatic argument unit identification (Achakulvisut et al., 2019; Li et al., 2021), automatic argument structure identification (Accuosto & Saggion, 2020; N. Song et al., 2019), and application, for example, evidence extraction (Li et al., 2021) or comparing claims made in articles (Yu et al., 2020). In another survey, the authors clarified that argumentative features can enhance NLP (Natural Language Processing) tasks (Lytos et al., 2019). They also proposed a theoretical framework which is an argumentation mining scheme, revealing the need for adopting more flexible and extensible frameworks.

When identifying argumentative discourse roles, especially argumentative zones, many of these studies follow the well-known argumentation model proposed by Teufel (1999). Teufel firstly defined this task and proposed the annotation scheme of 7 categories: Aim, Background, Basis, Contrast, Other, Own, Textual. In 2009, Teufel et al. (2009) proposed a new annotation scheme with 15 categories, Argumentative Zoning II (AZ-II), which is an elaboration of the original AZ scheme. Following these schemes, several corpora in other domains have been constructed, such as biomedical abstracts (Guo et al., 2011), papers in chemistry and computational linguistics (A. Yang & Li, 2018). Elizalde et al. (2016) generated a multi-layered annotated corpus of scientific discourse. Special features of the scientific discourse such as advantages and disadvantages are identified. Lauscher et al. (2018) present ArguminSci a tool that aims to support the holistic analyses of scientific publications in terms of scitorics, including the identification of argumentative components.

Except corpus construction, models can be learned for more scientific text mining tasks. Duma et al. (2016b) believed that query type is related to the type of citing sentences. In their work, the citing sentence is regarded as the user query and the citation context is the cited sentence. After annotating the citation context and citing sentences with CoreSC, their model learned the weight relationship of different types of user queries. Accuosto and Saggion (2020) explored two transfer learning approaches in which discourse parsing is used as an auxiliary task when training argument mining models. Song et al. (2020) adapt self-attention to discourse level for modeling discourse elements in argumentative student essays. Donkers and Ziegler (2020) made utilizations of argumentative information in user reviews for generating and explaining recommendations. Abbas et al. (2024) proposes a citation recommendation system using deep learning models to classify rhetorical zones of the research articles and compute similarity using rhetorical zone embeddings that overcome the cold-start problem. Chang et al. (2023) proposed a paper reading tool that leverages a user's publishing, reading, and saving activities to provide personalized visual augmentations and context around citations.

As we can see, the proposed argumentative categories will have some slight adjustments by different researchers to fit their own datasets (Kunnath et al., 2021). The new categories can then play a role in the corresponding models. Since there are very few attempts in citation recommendation task, we will propose a new classification schema and feed citing sentences with argumentative labels into our recommender network.

### 2.3. Citation Function in Citation Analysis

Scientific papers aim to present verifiable evidence for the stated claims. When citing paper, the authors also want to prove the relevance, validity and novelty of his/her main claims and conclusions (Pelclová & Lu, 2018). Many researchers are paying attention to citation analysis to help doing relevant text mining tasks (Y. K. Jeong et al., 2014; Kim et al., 2016; C. Zhang et al., 2021). Besides, citation sentences also contain some specific argumentative structure which can be utilized. For studies in citation analysis, such semantic information is referred to citation function or citation intent (Cohan et al., 2019).

Early in 1977, Spiegel-Rosing (1977) analyzed the uses of cited research in Science Studies. It is found that most frequent kind of use of cited research is to substantiate a statement or an assumption made in the citing text, or to point out further relevant information. Later, Teufel et al. (2006) conducted automatic classification of citation function based on his analysis. Abu-Jbara, Ezra and Radev (2013) adapted conclusion and schemes proposed by Spiegel-Rosing and Teufel et al. They used a taxonomy that consists of six categories, which are Criticizing, Comparison, Use, Substantiating, Basis and Neutral (Other). Jurgens et al. (2018) then performed the largest behavioral study of citations at that time, analyzing how scientific works frame their contributions through different types of citations and how this framing affects the field as a whole. Roman et al. (2021) used word embedding techniques to learn citation context vectors, then they create clusters of the word embeddings and assign a citation intent to each cluster. Maheshwari et al. (2021) finetuned BERT, RoBERTa, and SciBERT on training data to identify the purpose of citation sentences. Qi et al. (2023) proposed a multi-task learning model for citation intent classification by considering the correlation between citation intents, citation section and citation worthiness classification tasks. Budi and Yaniasih (2023) also proposed a multi-output model to analyze three citation meanings: sentiment, role and function. There are also some citation analysis tasks about citation recommendations. Cohan et al. (2019) regarded the task of identifying citation value (judging whether a citation sentence needs to be cited) as a subtask, and combined it with the function classification of citation sentences using a deep learning model. They improved the effect of citation sentence function classification through multi-task learning. Zhang and Zhu (2022) designed four forms of citation context based on citation motivations, then they extracted citation context and citation relationships to generate citation recommendations.

Currently, only a few works have utilized citation function information in citation recommendation task. More attempts should be done to figure out if it can help to improve the performance.

## 3. Methodology

To recommend citations based on argumentative zoning information of queries, there are two main steps to build the model. In the first step, we construct a corpus of labeled queries according to their argumentative zoning types. We applied citing sentences as user queries in this paper. The second step is to build a multi-task learning model which conduct citation recommendation and query classification at the same time. In this section, we will firstly give a brief introduction of Multi-Layer Perceptron (MLP), which is the basic neural network we applied in this paper. Then, the way that how we construct our corpus is explained. Finally, we will describe the details of our citation recommendation model and introduce our evaluation metrics and parameters.

## 3.1. Basics of Multi-Layer Perceptron Model

Inspired by the function of the brain, artificial neural networks have been found to be outstanding computing systems which can be applied in many disciplines. As one of the popular artificial neural networks, Multi-Layer Perceptron (MLP) applies a supervised training procedure using examples of data with known outputs (Bishop, 1995). To understand MLP, we will firstly introduce the single-layer perceptron. Single-layer perceptron only consist of one layer of input neurons and one layer of output neurons. Between the input and output, there is a single layer of weights. When looking into one perception in Figure 2, it has four parts[1]: the input layer and output layer, weights and bias, net sum, and the activation function. The input layer of the perceptron takes the initial data into the system for further processing. Weights represent the dimensions or strength of the connection between units. Bias is an additional parameter which is to modify the output along with the weighted sum of the input to the other neuron. Net sum calculates the total sum of the weights with inputs and the bias. The activation function calculates a weighted sum and further adding bias with it to give the result.

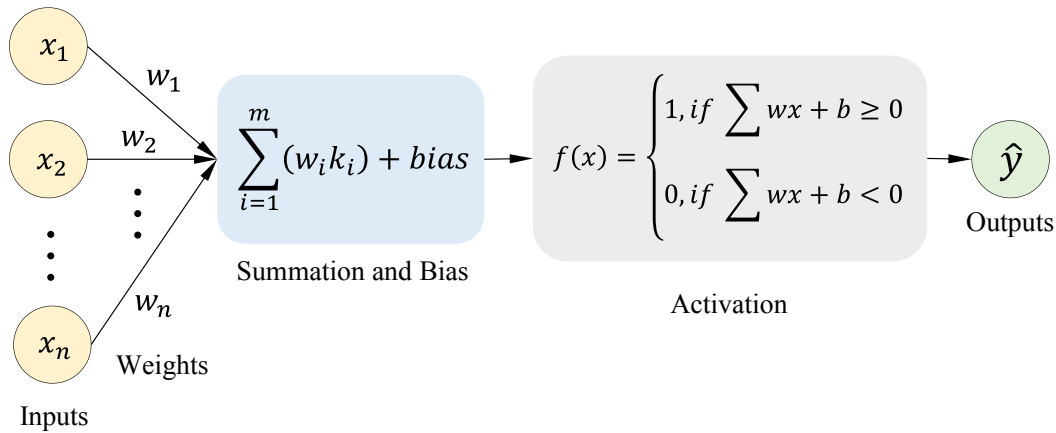

Figure 2. Structure of one perceptron

For the single-layer perceptron, it contains one or more perceptron in the network, shown in Figure 3(a). Since the single-layer perceptron is not able to figure out the nonlinearity or complexity of data. Researchers developed the multi-layer perceptron using the idea of the single-layer perceptron. The first and the last layers are input and output layers respectively, while the others are the hidden layers of the neural network (Taud & Mas, 2018). Figure 3(a) shows a single layer with three inputs and four outputs. Figure 3(b) shows the multi-layer with three inputs, two hidden layers and two outputs.

---

[1] Available at: https://www.javatpoint.com/single-layer-perceptron-in-tensorflow

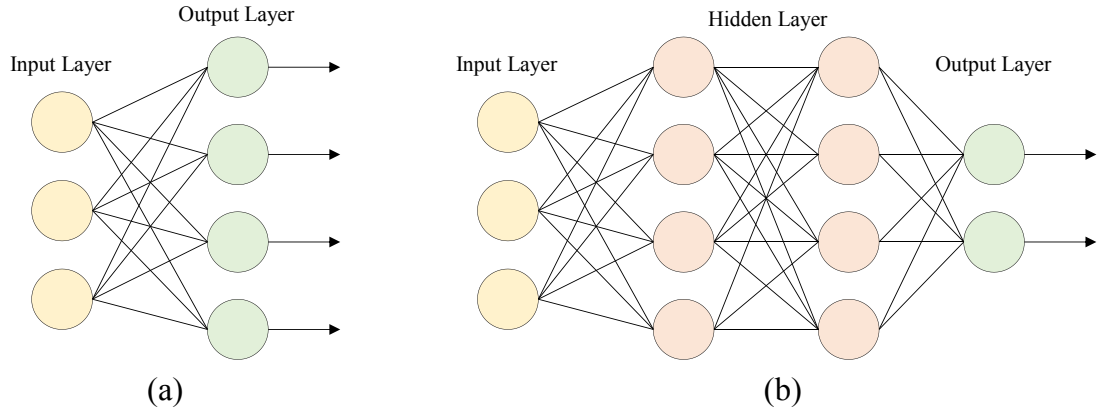

(a)                  (b)

**Figure 3. Structure of single layer perceptron**

The MLP is a layered feedforward neural network in which the information flows unidirectionally from the input layer to the output layer, passing through the hidden layers (Bishop, 1995). Each connection between neurons has its own weight. Perceptron for the same layer have the same activation function, which is used to determine the output of neural network.

### 3.2. Corpus Construction of User Queries

To better construct the dataset, we have adopted argumentative zoning taxonomy of user queries based on previous research. According to definition of different categories, guidelines for human annotation of basic scheme are then made. We will introduce the annotation instruction and dataset briefly. Detailed information will can be found in the website link of footnote. Finally, kappa coefficient is computed over the sample data to evaluate the agreement between annotators.

### 3.2.1. Classification of user queries

When setting the classification types, research related with argumentative zoning and citation function are referred. Basically, argumentative zoning task is to conduct sentence-by-sentence classification over scientific texts. Argumentative Zoning (Teufel, 1999), Argumentative Zoning II (Teufel et al., 2009) and Core Scientific Concepts (Duma et al., 2016a) are the three main schemas applied and modified in the most relevant studies. As Teufel et al. (2009) pointed out: parts of AZ scheme are similar to citation function classification. Normally, citation function reflects the specific purpose a citation plays with respect to the current paper's contributions (Jurgens et al., 2018). In our study, the aim is to improve the performance of citation recommendation by considering rhetorical information of user queries. Since our user query is citing sentences, we designed our schema based on the studies in argumentative zoning and citation function classification. The proposed categories, descriptions, and examples for each are listed in Table 1.

**Table 1. Classification of User Queries in our Dataset**

| Category | Description |
|---|---|

| Method | A description of the methods used in the research process. The methods here can refer to information such as experimental methods, procedures, data, information resources, tools, parameters, formats, standards, protocols, models, etc. |
|---|---|
| Conclusion | Conclusions based on experimental results and experimental phenomena. |
| Goal | A description of the research goal or an introduction to the research. |
| Object | Explanation or introduction of a research object or topic. |

### 3.2.2. Annotation guideline and dataset

The scientific texts we used are obtained from the PubMed Central database[2]. It is a free full-text archive of biomedical and life sciences journal literature at the U.S. National Institutes of Health's National Library of Medicine. There are two doctoral students with expertise in NLP for annotation. One of the annotators completed three rounds of pre-annotation to design guidelines and finish annotating the whole dataset. Another annotator followed the guideline and annotated 1000 pieces of user queries which are randomly sampled. Finally, we used the Kappa coefficient (Cohen, 1968) to measure the agreement.

Referring to the annotation guideline, for each query to be labeled, we will provide the sentence ID, original text, and its context information to the annotators. Besides, the citation location information is also supported. During the data processing, the citation mark is replaced with the string *[CITE]*. Therefore, annotators could consider the citation context information when labeling. One piece of data example is listed in Table 2. ID is the cited paper id, and original text is the citing sentence. Context information shows the raw data from PubMed which also contains the surrounding sentences (the paragraph that the citing sentence exist in).

**Table 2. Examples of User Queries to be Annotated**

| ID | Original Text | Context Information |
|---|---|---|
| b35 | Another form of granules important for RNA turnover are PBs, which can interact with SGs [CITE]. | Another form of granules important for RNA turnover are PBs, which can interact with SGs (<xref ref-type="bibr" rid="b35">35</xref>). Co-transfection of RFP-tagged PB marker DCP1a with GFP-tagged ZBP1, revealed an association with some DCP1a granules with full length ZBP1 or with the large granules formed by ZBP1;Z. |

In the annotation guidelines, we give the specific labeling rules for each category. Annotators can make judgement based on clue words and sentence patterns. The clue words refer to words that can bring clues to the identification of clear categories. For example, in the citing sentences below, the noun *method*, the verb *performed*, and the phrase *Spearman's rank correlation analyses* can all be used as clue words to distinguish the citation sentence as a *Method* category.

*As a method verification Spearman's rank correlation analyses of XER and XERcomp against determined blood levels of estradiol and testosterone (total and free) [CITE] were performed on the combined study group data.*

The clue words given for each category in the guidelines are not comprehensive. Therefore,

---
[2] Available at: https://www.ncbi.nlm.nih.gov/pmc/

we also provide some sentence template for annotators. For example, for *Goal* category, the sentence pattern template we provide is *someone do/does/did something*. The following sentence meets the template and is therefore judged as the *Goal* category:

*For instance, Masuda and colleagues [CITE] compared the molecular profile of the same RA FLSs cultured at low density (proliferating) and high density (quiescent).*

Normally, a user query can be only annotated with one category. When the sentence covers multiple categories and cannot be judged based on the citation location information and annotation guidelines, it will not be labeled. In this experiment, we also set a category priority to deal with such conflicts. Annotators can follow a certain order during labeling, that is, firstly label the whole data according to one specific category, and then label the remaining data with the remaining categories one by one. The category labeling order is *Method*, *Conclusion*, *Goal*, and *Object*. The complete version of annotation guideline is given in GitHub[3]. More comprehensive description and examples for different categories can be found there.

### 3.3. Citation Recommendation with Argumentative Zoning of User Query

Our citation recommendation model contains two main tasks: one is to predict the probability of candidate paper being cited; another one is to predict the argumentative category of user query. Previous studies have shown that the multi-task learning can exploit useful information between multiple tasks to help improving learning performance and exhibits promising results on many natural language processing tasks (Qi et al., 2023; Samant et al., 2022; Zhang & Yang, 2021). To verify the improvement of such multi-task model, we trained a Multilayer Perceptron (MLP) model only for citation recommendation (single task) to be the baseline. We will describe the network structures of multi-task model and single-task model. Evaluation metrics and parameters of model setting are also given in this section.

#### 3.3.1. Network structure of multi-task model

Figure 4 displays the structure of our proposed model which conducts multi-task learning. As it is shown, the input layer is the embeddings of user query and candidate paper (title and abstract). The word embedding size is set to be 200. After learning text features with the BiLSTM and attention structure (dimension of hidden layer and attention is set to be 128), we obtained vectors $x_{quey}, x_{title}, x_{abstract}$.

---

[3] Available at: https://github.com/michellemashutian/dissertation_citation_recommendation/blob/master/chapter5/annotation%20guideline.docx

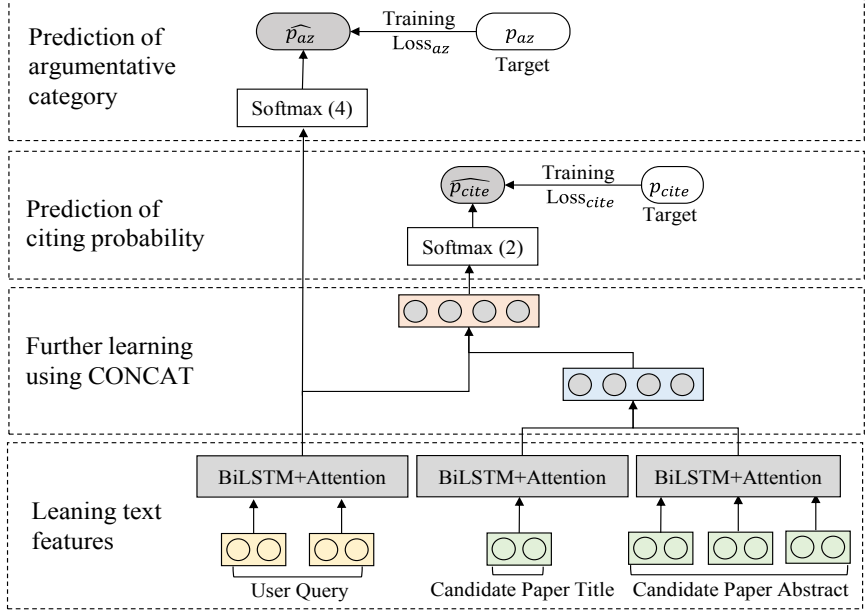

Figure 4. Network structure of the multi-task model

Then, in the first layer $L_1$, we try to combine the features of paper title and abstract by using CONCAT action. Dimension of layer $L_1$ is set to be 256. In the second layer $L_2$, we further combine the query vector with $L_1$. Dimension of layer $L_2$ is set to be 256. These two layers are defined as following:

$$L_1 = ReLU(W_1[x_{title}, x_{abstract}] + b_1)$$

$$L_2 = ReLU(W_2[x_{query}, L_1] + b_2)$$

$W_1, W_2$ and $b_1, b_2$ denote the related weights and biases, respectively. The Rectified Linear Unit (ReLU) is applied as activation function to generate the output of a neuron (Empirical evaluation of rectified activations in convolutional network). Formula of ReLU is given below:

$$f(x) = max\,(0, x)$$

For the task of citation recommendation, probability to cite the candidate paper $\widehat{P_{cite}}$ is achieved by applying a softmax function over the last fully connected layer:

$$\widehat{P_{cite}} = softmax(W_3 L_2 + b_3)$$

$W_3$ and $b_3$ denote the related weight and bias, respectively. Suppose for vector $V$, and $V_i$ is the ith element, softmax value of $V_i$ can be calculated by following formula:

$$S_i = \frac{e^{V_i}}{\sum_j e^{V_j}}$$

For the task of argumentative zoning for user query, we also use SoftMax function over a fully connected layer to obtain the probability distribution of different categories. Dimension of the fully connected layer is set to be 20. $\widehat{P_{az}}$ is computed as follows:

$$\widehat{P_{az}} = softmax(W_{az} x_{citation} + b_{az})$$

$W_{az}$ and $b_{az}$ denote the related weight and bias, respectively. During model training, cross

entropy between true probability $p$ and predicted probability $\hat{p}$ is chosen to be loss function and its formula is given below:

$$Loss(p, \hat{p}) = -\sum_{i=0}^{|p|} p_i log\hat{p_i}$$

The losses from two different prediction tasks are then combined with weights to be the total loss to optimize. To test different parameter settings, $\alpha$ is set to be 0.1, 0.2 and 0.3.

$$Loss = Loss_{citing} + \alpha * Loss_{az}$$

Adam algorithm (Kingma & Ba, 2014) is applied for efficient stochastic optimization and learning rate is set to be 0.001. In this experiment, we used the trained model to predict probabilities of citing and not citing the candidate paper in testing data. If the citing probability is higher, then this paper will be recommended. Probability distribution over different argumentative categories is also obtained. We choose the category with highest probability as the one to which the candidate paper belongs.

### 3.3.2. Network structure of single-task model

Figure 5 displays the structure of the baseline model which only conducts citation recommendation. Long Short-Term Memory networks (LSTMs) are a widely used type of recurrent neural network and are often regarded as a standard baseline for deep learning models (Chen, Wang, & Wang, 2023; Melis, Dyer, & Blunsom, 2017; Usmani & Shamsi, 2023). One such variant, the Bi-directional LSTM (BiLSTM), enhances the original LSTM by processing input sequences in both forward and backward directions (Hameed & Garcia-Zapirain, 2020; Li, Fu, & Ma, 2020; Tzoumpas, Estrada, Miraglio, & Zambelli, 2024). In our baseline setting, we applied BiLSTM and attention structure to learn text features of user query, candidate paper title and abstract.

After two fully connected layers, we combined all textual inputs and applied a softmax function over the last fully connected layer to predict the probability of citing and not citing the candidate paper. These layers are defined as following:

Layer1:

$$L_1 = ReLU(W_1[x_{title}, x_{abstract}] + b_1)$$

Layer2:

$$L_2 = ReLU(W_2[x_{query}, L_1] + b_2)$$

Output Layer:

$$\hat{p} = softmax(W_3 L_2 + b_3)$$

$W_1, W_2, W_3$ and $b_1, b_2, b_3$ denote the related weights and biases, respectively. Activation function is ReLU. Cross entropy between true probability $p$ and predicted probability $\hat{p}$ is chosen to be loss function. Adam optimization algorithm is used for stochastic gradient descent for training model and learning rate is set to be 0.001. Finally, the algorithm will give recommendations based on the probability of citing and not citing.

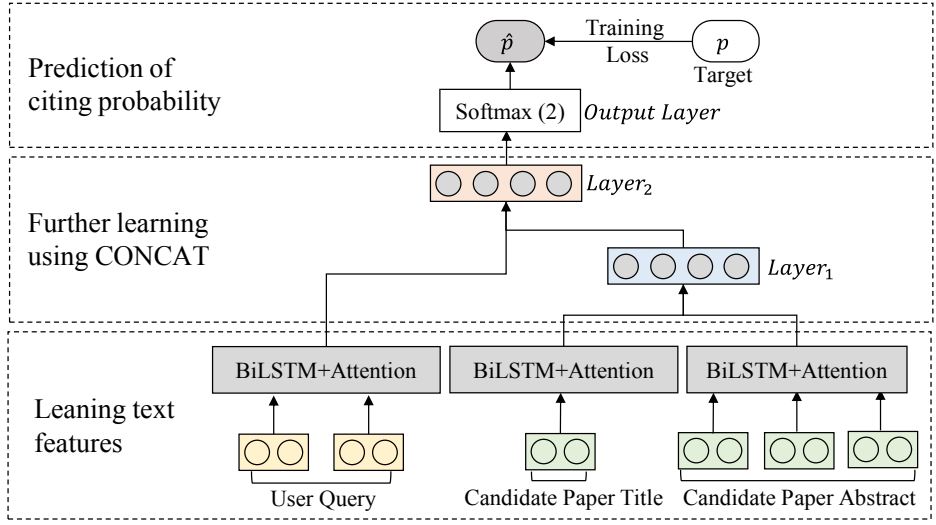

**Figure 5. Network structure of the single-task model**

### 3.3.3. Evaluation metrics and parameters

Finally, the algorithm will decide whether to recommend based on the higher probability among the probabilities of citing and not citing. When evaluating models of citing paper prediction and not citing paper prediction, we applied macro value of precision (Macro_P), recall (Marco_R) and $F_1$ (Marco_$F_1$). Formulas are given as follows:

$$Macro\_P = \frac{1}{n}\sum_{1}^{n} P_i$$

$$Macro\_R = \frac{1}{n}\sum_{1}^{n} R_i$$

$$Macro\_F_1 = \frac{1}{n}\sum_{1}^{n} F_{1i}$$

Furthermore, classification of argumentative category is evaluated based precision, recall and $F_1$. Formulas are given as follows:

$$P = \frac{TP}{TP + FP}$$

$$R = \frac{TP}{TP + FN}$$

$$F_1 = \frac{2 \times P \times R}{P + R}$$

Where, $TP$ is the number of true positive data, $FP$ is the false positive data and $FN$ is the false negative data.

## 4. Experiment and Result Analysis

### 4.1. Experimental Setup

This section will introduce the experimental data, evaluation metrics, model parameter setting, and the baselines applied in this paper.

#### 4.1.1. Experimental data

In this paper, we applied the dataset from PubMed Central database[4]. 544, 511 publications between 1877 to 2013 with 916, 860 citation relations are extracted. The corpus is divided into seven time slices shown in Table 3.

Table 3. Paper numbers of different time slices of PubMed dataset

| Time Slice | Paper Number | Time Slice | Paper Number |
|---|---|---|---|
| Pre-1995 | 68, 523 | 2006-2007 | 82, 868 |
| 1996-2000 | 77, 290 | 2008-2009 | 90, 165 |
| 2001-2003 | 77, 504 | 2010-2013 | 77, 954 |
| 2004-2005 | 70, 207 | - | - |

To select the citing sentences with different argumentative information, we filter papers to keep those which have over 30 references and their references are published in more than 5 time slices. Then, there are 12, 882 papers left. When downloading the papers from file transfer protocol of National Center for Biotechnology Information[5], we extract the information of papers ID, title, abstract, full text, and references ID. Based on the 12, 882 papers and their references, we found 40, 758 pairs of citing paper and cited paper where the cited paper has full-text data. Since it costs a lot of time and labor for data annotation. We randomly select 15, 000 pairs of citing sentence and cited paper to label[6]. Among them, only 9217 pieces of citing sentences can be classified into a certain argumentative category. For the left ones, we mark them as "*Other*" category. Then, these data will be used as our experimental dataset. The paper numbers of different categories are shown in Table 4.

Table 4. Paper numbers of different argumentative categories in original dataset

| Argumentative Category | Paper Number | Percentage |
|---|---|---|
| Method | 3859 | 0.257 |
| Conclusion | 3144 | 0.210 |
| Goal | 294 | 0.020 |
| Object | 1920 | 0.128 |
| Other | 5783 | 0.385 |

---

[4] Available at: https://pubmed.ncbi.nlm.nih.gov/

[5] Available at: ftp://ftp.ncbi.nlm.nih.gov/pub/pmc/oa_bulk/

[6] Available at: https://github.com/michellemashutian/dissertation_citation_recommendation/blob/master/chapter5/file5.xlsx

For the 9217 pairs of citing sentence and cited paper, we parsed the XML data based on the paper ID and obtained the information of citing paper, cited paper title and abstract, publication time of cited paper. Among all the dataset, there are 127 papers having no abstract information. Then, we use the very first paragraph to represent this paper. To construct a certain ratio of positive and negative sample data, we set the positive and negative ratio to be 1:5. That is, for each citing sentence/user query, if the number of citations for the query is $k$, we will construct $5k$ negative sample samples and their data set.

The negative examples construction process is as follows: firstly, we calculate the similarities between user query and other candidate papers. Then, negative samples are selected based on three similarity criteria: papers with highest similarity to the user query, papers with the lowest similarity to the user query, and papers which are in the median position of all similarities. The sampling ratio of these three types of papers is 5:2:3. In the last step, we integrate the negative and positive data as the final dataset. Finally, there are 55, 302 pairs of user query and candidate citation with information of citing or not and argumentative category in total. We randomly select 5000 pieces with specific argumentative information from the final dataset to be the testing data and the others are used as training one. For this testing data, paper numbers of argumentative categories are displayed in Table 5.

Table 5. Paper numbers of different argumentative categories in testing data

| Argumentative Category | Paper Number | Percentage |
| --- | --- | --- |
| Method | 2076 | 0.415 |
| Conclusion | 1731 | 0.346 |
| Goal | 163 | 0.033 |
| Object | 1030 | 0.206 |

**4.1.2. Annotation agreement**

To check the inter-annotator agreement, we randomly picked 1000 sentences from the whole dataset and let two annotators to do the labeling work. We use Kappa coefficient (Cohen, 1968) to measure the agreement. Suppose there are Annotator 1 and Annotator 2 labeling data over Category A and Category B. The annotation results are shown as Table 6, where *a* means the number of data that both annotators labeled with Category A, *b* means the number of data that Annotator 1 labeled with Category B, but Annotator 2 labeled with Category A.

Table 6. Annotation Results of Annotator 1 and Annotator 2

| Results from Annotator 2 | Results from Annotator 1 | |
| --- | --- | --- |
| | Category A | Category B |
| **Category A** | a | b |
| **Category B** | c | d |

So, the formula for Kappa coefficient is shown as follows:

$$Kappa = \frac{P_o - P_e}{1 - P_e}$$

Where, $P_o$ is the relative observed agreement among annotators and $P_e$ is the hypothetical

probability of chance agreement. Suppose $n = a + b + c + d$, $P_o$ and $P_e$ are computed as follows:

$$P_o = \frac{a + d}{n}$$

$$P_e = \frac{(a + b)(a + c) + (c + d)(b + d)}{n^2}$$

Real annotation results by the two annotators in this experiment is shown in **Table 7**. Previous research has shown that: when Kappa is over 0.6, this value indicates almost perfect agreement. When Kappa is between 0.4 to 0.6, this value indicates substantial agreement. When Kappa is below 0.4, this value indicates poor agreement. The Kappa coefficient obtained by our two annotators is 0.6819, which shows substantial agreement. So, we can conduct further research over the labeled dataset.

Table 7. Annotation Results of two Annotators in this Experiment

| Results from Annotator 2 | Results from Annotator 1 | | | |
|---|---|---|---|---|
| | Method | Conclusion | Goal | Object |
| Method | 347 | 35 | 7 | 23 |
| Conclusion | 13 | 285 | 6 | 52 |
| Goal | 7 | 11 | 16 | 1 |
| Object | 16 | 40 | 1 | 140 |

### 4.2. Result Analysis

In this section, we analyze the experimental results from two aspects: performance of citation recommendation and performance of argumentative classification task.

#### 4.2.1. Performance of citation recommendation

Table 8 shows macro evaluation metrics of citation recommendation. As we can see, with considering the argumentative information, when $\alpha = 0.1$ and $\alpha = 0.2$, multi-task model shows better performance than the baseline model which applied the single-task structure Besides, the parameter $\alpha$ of loss weight is important to set to obtain a better performance. By sharing the representations of user query and candidate papers between tasks of citation recommendation and argumentative zoning, we can enable our model to generalize better on our original task.

Table 8. Macro evaluation metrics of different models

| Models \ Metrics | Macro_P | Marco_R | Marcro_F$_1$ |
|---|---|---|---|
| Single-task model (baseline) | 0.7908 | 0.5644 | 0.5743 |
| Multi-task model ($\alpha = 0.1$) | 0.7270 | 0.6123 | ***0.6370*** |
| Multi-task model ($\alpha = 0.2$) | 0.7648 | 0.5720 | 0.5860 |
| Multi-task model ($\alpha = 0.3$) | 0.7681 | 0.5637 | 0.5733 |

Table 9 shows evaluation metrics of papers which should be cited (positive samples) and papers which should not be cited (negative samples) separately. From Table 9, we can find that, after considering the argumentative information of user query, although precision of positive samples is getting lower, there is a great improvement for recall. It reflects that argumentative category of user query is helpful for identifying the true citations. The multi-task learning model significantly outperforms the baseline single-task model in overall performance, particularly in the citation recommendation process. By accurately recognizing the argumentative category, the model can more effectively assess and recommend relevant citations for the query.

**Table 9. Evaluation metrics of positive and negative samples**

| Metrics / Models | Positive Samples | | | Negative Samples | | |
| --- | --- | --- | --- | --- | --- | --- |
| | P | R | $F_1$ | P | R | $F_1$ |
| Single-task model (baseline) | 0.7313 | 0.1391 | 0.2338 | 0.8504 | 0.9897 | 0.9148 |
| Multi-task model ($\alpha = 0.1$) | 0.5882 | 0.2616 | 0.3621 | 0.8658 | 0.9630 | 0.9118 |
| Multi-task model ($\alpha = 0.2$) | 0.6768 | 0.1593 | 0.2579 | 0.8528 | 0.9846 | 0.9140 |
| Multi-task model ($\alpha = 0.3$) | 0.6861 | 0.1403 | 0.2330 | 0.8503 | 0.9870 | 0.9135 |

#### 4.2.2. Performance of argumentative classification

Table 10 shows macro evaluation metrics of argumentative classification. Table 11 shows evaluation metrics of different categories separately. From Table 10 and Table 11, we can find that, with the increase of parameter $\alpha$, performance of argumentative classification is getting better since the model can optimize more loss derived from this task. Among all different categories, when $\alpha = 0.3$, the model performs the best compared with the other two settings. Besides, when judging the *Method* and *Conclusion* category, the accuracy, recall and $F_1$ value of each model is all higher than the other two categories. It shows that our proposed model has a better recognition performance over *Method* and *Conclusion* categories. According to the data distribution of different categories given in Table 4, the imbalanced dataset might be the reason for such phenomena. So, in the future work, improvement can be firstly done by extending the training dataset of categories with little data volume.

**Table 10. Macro evaluation metrics of argumentative classification**

| Metrics / Models | Macro_P | Marco_R | Marcro_$F_1$ |
| --- | --- | --- | --- |
| Multi-task model ($\alpha = 0.1$) | 0.9323 | 0.8656 | 0.8914 |
| Multi-task model ($\alpha = 0.2$) | 0.9652 | 0.9289 | 0.9453 |
| Multi-task model ($\alpha = 0.3$) | 0.9806 | 0.9554 | ***0.9671*** |

**Table 11. Evaluation metrics of different argumentative categories**

| Metrics / Models | Method | | | Conclusion | | |
| --- | --- | --- | --- | --- | --- | --- |
| | P | R | $F_1$ | P | R | $F_1$ |
| Multi-task model ($\alpha = 0.1$) | 0.9592 | 0.9620 | 0.9606 | 0.9344 | 0.9301 | 0.9323 |
| Multi-task model ($\alpha = 0.2$) | 0.9882 | 0.9697 | 0.9789 | 0.9258 | 0.9879 | 0.9558 |
| Multi-task model ($\alpha = 0.3$) | 0.9918 | 0.9841 | 0.9879 | 0.9721 | 0.9861 | 0.9791 |
| Metrics / Models | Goal | | | Object | | |
| | P | R | $F_1$ | R | P | F1 |
| Multi-task model ($\alpha = 0.1$) | 0.9730 | 0.6626 | 0.7883 | 0.8626 | 0.9078 | 0.8846 |
| Multi-task model ($\alpha = 0.2$) | 0.9786 | 0.8405 | 0.9043 | 0.9682 | 0.9175 | 0.9422 |
| Multi-task model ($\alpha = 0.3$) | 1.0000 | 0.8834 | 0.9381 | 0.9587 | 0.9680 | 0.9633 |

### 4.2.3. Analysis of recommended papers

From the previous analysis, we can find that there is an increase of recall value when adding the task of argumentative zoning. To clearly see the difference of recall changes between single-task model and multi-task model, we count the number of samples that have been recalled among different categories in Table 12.

**Table 12. Paper numbers that have been recalled by different models**

| Category \ Numbers | Papers only recalled by multi-task model | Papers only recalled by single-task model |
|---|---|---|
| Method | 17 | 9 |
| Conclusion | 13 | 4 |
| Goal | 1 | 0 |
| Object | 7 | 5 |

From Table 12, it can be seen that most of the citations recalled by the multi-task model belong to the *Method* category and the *Conclusion* category. We further display two examples of citing probability predicted by single-task and multi-task models.

***User Query 1***: *Such effective networks are a novel and highly instructive way of exploring the relation between network architecture and dynamical processes (see for an analysis of effective gene regulatory networks and [CITE], for a theoretical study of effective networks).*

For user query 1, the citing probiliaty predicted by single-task model is 42.19%. However, there is a increase in multi-task model, citing probability is 67.54% and the argumentative classification module predicts it as the *Method* category. As we can see, this query aims to explore ways to explore the relationship between network structure and dynamic processes, and when we further look over the actual citation (ID: 19826610), it is a paper titled with *Interplay Between Topology and Dynamics in Excitation Patterns on Hierarchical Graphs*[7], from the abstract of this paper, we can find the discussion about the topology of the graph and the dynamics of the network:

*In a recent publication (Müller-Linow et al., 2008) two types of correlations between network topology and dynamics have been observed: waves propagating from central nodes and module-based synchronization. Remarkably, the dynamic behavior of hierarchical modular networks can switch from one of these modes to the other as the level of spontaneous network activation changes. Here we attempt to capture the origin of this switching behavior in a mean-field model as well in a formalism, where excitation waves are regarded as avalanches on the graph.*

---

[7] Available at: https://pubmed.ncbi.nlm.nih.gov/19826610/

*User Query 2: An early study reported CCRL2 binding of CCL5 as well as several CCR2-ligands including CCL2, with resulting signaling and cell migration, which, however, could not be independently confirmed [CITE].*

For user query 2, the citing probiliaty predicted by single-task model is 49.51%. In multi-task model, citing probability is 56.28% and the argumentative classification module predicts it as the *Conclusion* category. The actual citation (ID: 19826610) is a paper titled with *Mast cell-expressed orphan receptor CCRL2 binds chemerin and is required for optimal induction of IgE-mediated passive cutaneous anaphylaxis*[8], from the abstract of this paper, we can see the authors conducted research to show finding that the mast cell-expressed orphan serpentine receptor mCCRL2 is not required for expression of IgE-mediated mast cell-dependent passive cutaneous anaphylaxis:

*Mast cells contribute importantly to both protective and pathological IgE-dependent immune responses. We show that the mast cell-expressed orphan serpentine receptor mCCRL2 is not required for expression of IgE-mediated mast cell-dependent passive cutaneous anaphylaxis but can enhance the tissue swelling and leukocyte infiltrates associated with such reactions in mice. We further identify chemerin as a natural nonsignaling protein ligand for both human and mouse CCRL2. In contrast to other "silent" or professional chemokine interreceptors, chemerin binding does not trigger ligand internalization. Rather, CCRL2 is able to bind the chemoattractant and increase local concentrations of bioactive chemerin, thus providing a link between CCRL2 expression and inflammation via the cell-signaling chemerin receptor CMKLR1.*

These two examples belong to a clear type of argumentation category. After argumentative classification is added to the neural network model, the citation probability output by the model is significantly improved, thereby improving the recall metric. To sum up, although in single-task model, the neural network cannot capture the semantic correlation between the user query and the candidate citation, but after considering the type of argument structure, the citation probability of the candidate citation is getting changed.

## 5. Conclusion and Future Work

This paper proposed a multi-task model that combining argumentative zoning of user query with citation recommendation task. Based on the experimental results, we can find that by adding argumentative information into model, performance of citation recommendation can be further improved. Besides, such rhetorical clues are helpful in other citation-related task for model refinement as well. It is flexible and convenient to add our argumentative zoning module in neural network. Before conducting the classification of user queries, this paper also builds a new argumentative taxonomy and performs a relatively expensive manual labeling work on PubMed dataset. A random set of data has passed the consistency test. Our annotation corpus is available to all researchers. The methodological limitations of this work are as follows. Our model requires a certain amount of well-labeled training corpus. If such model is trained on a larger corpus, more labor and time costs is required. Although the argumentative information of user queries will enhance the recommenders. We only list some basic argumentative categories in this paper. The actual application may be more complicated.

---

[8] Available at: https://pubmed.ncbi.nlm.nih.gov/18794339/

In the future, the attention can be paid in several aspects to optimize the model. The first is to expand our corpus and make data augmentation over the categories with few data amount. The second is to apply more features except argumentative categories, for example, the period of paper publication time or the topic word. The third is to conduct more experiments by adding the argumentative zoning module over other citation recommenders to see if it can also work in other models. Currently, with the burst of large language model, we can also perform fine-tuning of a large language model by using our labeled dataset to help enhance the recommender's performance.

## Acknowledgement

This study has received support from the National Natural Science Foundation of China (Grant No. 72074113) and the Major Project of the National Social Science Fund of China (Grant No. 20ZDA039).

## References

Abu-Jbara, A., Ezra, J., & Radev, D. (2013). Purpose and Polarity of Citation: Towards NLP-based Bibliometrics. Proceedings of the 2013 Conference of the North American Chapter of the Association for Computational Linguistics: Human Language Technologies, 596–606. https://aclanthology.org/N13-1067

Abbas, M. A., Ajayi, S., Bilal, M., Oyegoke, A., Pasha, M., & Ali, H. T. (2024). A deep learning approach for context-aware citation recommendation using rhetorical zone classification and similarity to overcome cold-start problem. Journal of Ambient Intelligence and Humanized Computing, 15(1), 419-433.

AbuRa'ed, A. G. T., Chiruzzo, L., & Saggion, H. (2018). Experiments in detection of implicit citations. Paper presented at the WOSP 2018. 7th International Workshop on Mining Scientific Publications; 2018 May 7; Miyazaki, Japan.[Paris (Francce)]: European Language Resources Association; 2018. 7 p.

Ali, Z., Ullah, I., Khan, A., Ullah Jan, A., & Muhammad, K. (2021). An overview and evaluation of citation recommendation models. Scientometrics, 126, 4083-4119.

Accuosto, P., & Saggion, H. (2020). Mining arguments in scientific abstracts with discourse-level embeddings. Data & Knowledge Engineering, 129, 101840. https://doi.org/10.1016/j.datak.2020.101840

Achakulvisut, T., Bhagavatula, C., Acuña, D. E., & Körding, K. P. (2019). Claim Extraction in Biomedical Publications using Deep Discourse Model and Transfer Learning. CoRR, abs/1907.00962. http://arxiv.org/abs/1907.00962

Al Khatib, K., Ghosal, T., Hou, Y., de Waard, A., & Freitag, D. (2021). Argument Mining for Scholarly Document Processing: Taking Stock and Looking Ahead. Proceedings of the Second Workshop on Scholarly Document Processing, 56–65. https://doi.org/10.18653/v1/2021.sdp-1.7

Alhoori, H., & Furuta, R. (2017). Recommendation of scholarly venues based on dynamic user interests. Journal of Informetrics, 11(2), 553–563. https://doi.org/10.1016/j.joi.2017.03.006

Bishop, C. M. (1995). Neural networks for pattern recognition: Oxford university press.

Budi, I., & Yaniasih, Y. (2023). Understanding the meanings of citations using sentiment, role, and citation function classifications. Scientometrics, 128(1), 735-759.

Chang, J. C., Zhang, A. X., Bragg, J., Head, A., Lo, K., Downey, D., & Weld, D. S. (2023). Citesee: Augmenting citations in scientific papers with persistent and personalized historical context. Paper presented at the Proceedings of the 2023 CHI Conference on Human Factors in Computing Systems.

Chen, Z., Wang, M., & Wang, Y. (2023). Improving indoor occupancy detection accuracy of the SLEEPIR sensor using LSTM models. IEEE Sensors Journal, 23(15), 17794-17802.

Cohan, A., Ammar, W., van Zuylen, M., & Cady, F. (2019). Structural Scaffolds for Citation Intent Classification in Scientific Publications. Proceedings of the 2019 Conference of the North American Chapter of the Association for Computational Linguistics: Human Language Technologies, Volume 1 (Long and Short Papers), 3586–3596. https://doi.org/10.18653/v1/N19-1361

Cohen, J. (1968). Weighted kappa: Nominal scale agreement provision for scaled disagreement or partial credit. Psychological Bulletin, 70(4), 213–220. https://doi.org/10.1037/h0026256

Dai, T., Zhu, L., Cai, X., Pan, S., & Yuan, S. (2018). Explore semantic topics and author communities for citation recommendation in bipartite bibliographic network. Journal of Ambient Intelligence and Humanized Computing, 9(4), 957–975. https://doi.org/10.1007/s12652-017-0497-1

Donkers, T., & Ziegler, J. (2020). Leveraging Arguments in User Reviews for Generating and Explaining Recommendations. Datenbank-Spektrum, 20(2), 181–187. https://doi.org/10.1007/s13222-020-00350-y

Duma, D., Liakata, M., Clare, A., Ravenscroft, J., & Klein, E. (2016a). Applying Core Scientific Concepts to Context-Based Citation Recommendation. Proceedings of the Tenth International Conference on Language Resources and Evaluation (LREC'16), 1737–1742. https://aclanthology.org/L16-1274

Duma, D., Liakata, M., Clare, A., Ravenscroft, J., & Klein, E. (2016b). Rhetorical Classification of Anchor Text for Citation Recommendation. D-Lib Magazine, 22(9/10). https://doi.org/10.1045/september2016-duma

Editors, I. C. of M. J. & others. (2004). Uniform requirements for manuscripts submitted to biomedical journals: Writing and editing for biomedical publication.

Färber, M., & Jatowt, A. (2020). Citation recommendation: approaches and datasets. International Journal on Digital Libraries, 21(4), 375-405.

Fisas Elizalde, B., Ronzano, F., & Saggion, H. (2016). A multi-layered annotated corpus of scientific papers. Paper presented at the Calzolari N, Choukri K, Declerck T, Goggi S, Grobelnik M, Maegaard B, Mariani J, Mazo H, Moreno A, Odijk J, Piperidis S, editors. LREC 2016. Tenth International Conference on Language Resources and Evaluation; 2016 May 23-28; Portorož, Slovenia.[Paris]: ELRA; 2016. p. 3081-8.

Guo, Y., Korhonen, A., & Poibeau, T. (2011). A Weakly-Supervised Approach to Argumentative Zoning of Scientific Documents. Proceedings of the Conference on Empirical Methods in Natural Language Processing, 273–283.

Gündoğan, E., Kaya, M., & Daud, A. (2023). Deep learning for journal recommendation system of research papers. Scientometrics, 128(1), 461-481.

He, Q., Pei, J., Kifer, D., Mitra, P., & Giles, L. (2010). Context-aware citation recommendation. Proceedings of the 19th International Conference on World Wide Web, 421–430. https://doi.org/10.1145/1772690.1772734

Hua, X., Nikolov, M., Badugu, N., & Wang, L. (2019). Argument Mining for Understanding Peer Reviews. Proceedings of the 2019 Conference of the North American Chapter of the Association for Computational Linguistics: Human Language Technologies, Volume 1 (Long and Short Papers), 2131–2137. https://doi.org/10.18653/v1/N19-1219

Kunnath, S. N., Herrmannova, D., Pride, D., & Knoth, P. (2021). A meta-analysis of semantic classification of citations. Quantitative science studies, 2(4), 1170-1215.

Jeong, C., Jang, S., Park, E., & Choi, S. (2020). A context-aware citation recommendation model with BERT and graph convolutional networks. Scientometrics, 124(3), 1907–1922. https://doi.org/10.1007/s11192-020-03561-y

Jeong, Y. K., Song, M., & Ding, Y. (2014). Content-based author co-citation analysis. Journal of Informetrics, 8(1), 197–211. https://doi.org/10.1016/j.joi.2013.12.001

Jurgens, D., Kumar, S., Hoover, R., McFarland, D., & Jurafsky, D. (2018). Measuring the Evolution of a Scientific Field through Citation Frames. Transactions of the Association for Computational Linguistics, 6, 391–406. https://doi.org/10.1162/tacl_a_00028

Kim, H. J., Jeong, Y. K., & Song, M. (2016). Content- and proximity-based author co-citation analysis using citation sentences. Journal of Informetrics, 10(4), 954–966. https://doi.org/10.1016/j.joi.2016.07.007

Kingma, D. P., & Ba, J. (2014). Adam: A Method for Stochastic Optimization. https://arxiv.org/abs/1412.6980v9

Lauscher, A., Glavaš, G., & Ponzetto, S. P. (2018). An Argument-Annotated Corpus of Scientific Publications. Proceedings of the 5th Workshop on Argument Mining, 40–46. https://doi.org/10.18653/v1/W18-5206

Lauscher, A., Glavaš, G., & Eckert, K. (2018). ArguminSci: A tool for analyzing argumentation and rhetorical aspects in scientific writing. Paper presented at the Proceedings of the 5th Workshop on Argument Mining.

Li, P.-H., Fu, T.-J., & Ma, W.-Y. (2020). Why attention? Analyze BiLSTM deficiency and its remedies in the case of NER. Paper presented at the Proceedings of the AAAI conference on artificial intelligence.

Li, X., Burns, G. A., & Peng, N. (2021). Scientific Discourse Tagging for Evidence Extraction. In P. Merlo, J. Tiedemann, & R. Tsarfaty (Eds.), Proceedings of the 16th Conference of the European Chapter of the Association for Computational Linguistics: Main Volume, EACL 2021, Online, April 19—23, 2021 (pp. 2550–2562). Association for Computational Linguistics. https://aclanthology.org/2021.eacl-main.218/

Liakata, M., Teufel, S., Siddharthan, A., & Batchelor, C. (2010). Corpora for the conceptualisation and zoning of scientific papers. LREC 2010, 7th International Conference on Language Resources and Evaluation, Valletta, Malta. http://citeseerx.ist.psu.edu/viewdoc/download?doi=10.1.1.724.8203&rep=rep1&type=pdf

Liu, J.-C., Chen, C.-T., Lee, C., & Huang, S.-H. (2024). Evolving Knowledge Graph Representation Learning with Multiple Attention Strategies for Citation Recommendation System. ACM Transactions on Intelligent Systems and Technology, 15(2), 1-26.

Lytos, A., Lagkas, T., Sarigiannidis, P., & Bontcheva, K. (2019). The evolution of argumentation mining: From models to social media and emerging tools. Information Processing & Management, 56(6), 102055.

Ma, S., Zhang, C., & Liu, X. (2020). A review of citation recommendation: From textual content to enriched context. Scientometrics, 122(3), 1445–1472. https://doi.org/10.1007/s11192-019-03336-0

Ma, X., & Wang, R. (2019). Personalized Scientific Paper Recommendation Based on Heterogeneous Graph Representation. IEEE Access, 7, 79887–79894. https://doi.org/10.1109/ACCESS.2019.2923293

Maheshwari, H., Singh, B., & Varma, V. (2021). SciBERT Sentence Representation for Citation Context Classification. Proceedings of the Second Workshop on Scholarly Document Processing, 130–133. https://aclanthology.org/2021.sdp-1.17

Mei, X., Cai, X., Xu, S., Li, W., Pan, S., & Yang, L. (2022). Mutually reinforced network embedding: An integrated approach to research paper recommendation. Expert Systems with Applications, 204, 117616.

Melis, G., Dyer, C., & Blunsom, P. (2017). On the state of the art of evaluation in neural language models. arXiv preprint arXiv:1707.05589.

Muther, R., & Smith, D. (2023). Citations as Queries: Source Attribution Using Language Models as Rerankers. arXiv preprint arXiv:2306.17322.

Pelclová, J., & Lu, W. (Eds.). (2018). Persuasion in public discourse: Cognitive and functional perspectives. John Benjamins Publishing Company.

Pornprasit, C., Liu, X., Kiattipadungkul, P., Kertkeidkachorn, N., Kim, K.-S., Noraset, T., Hassan, S.-U., & Tuarob, S. (2022). Enhancing citation recommendation using citation network embedding. Scientometrics, 127(1), 233–264. https://doi.org/10.1007/s11192-021-04196-3

Qi, R., Wei, J., Shao, Z., Li, Z., Chen, H., Sun, Y., & Li, S. (2023). Multi-task learning model for citation intent classification in scientific publications. Scientometrics, 1-21.

Rollins, J., McCusker, M., Carlson, J., & Stroll, J. (2017). Manuscript Matcher: A Content and Bibliometrics-based Scholarly Journal Recommendation System. 12.

Roman, M., Shahid, A., Khan, S., Koubaa, A., & Yu, L. (2021). Citation Intent Classification Using Word Embedding. IEEE Access, 9, 9982–9995. https://doi.org/10.1109/ACCESS.2021.3050547

Samant, R. M., Bachute, M. R., Gite, S., & Kotecha, K. (2022). Framework for deep learning-based language models using multi-task learning in natural language understanding: A systematic literature review and future directions. IEEE Access, 10, 17078-17097.

Song, N., Cheng, H., Zhou, H., & Wang, X. (2019). Argument Structure Mining in Scientific Articles: A Comparative Analysis. Proceedings of the 18th Joint Conference on Digital Libraries, 339–340. https://doi.org/10.1109/JCDL.2019.00060

Song, W., Song, Z., Fu, R., Liu, L., Cheng, M., & Liu, T. (2020). Discourse Self-Attention for Discourse Element Identification in Argumentative Student Essays. Proceedings of the 2020 Conference on Empirical Methods in Natural Language Processing (EMNLP), 2820–2830. https://doi.org/10.18653/v1/2020.emnlp-main.225

Spiegel-Rosing, I. (1977). Science Studies: Bibliometric and Content Analysis. Social Studies of Science, 7(1), 97–113. https://doi.org/10.1177/030631277700700111

Tang, J., & Zhang, J. (2009). A Discriminative Approach to Topic-Based Citation Recommendation. In T. Theeramunkong, B. Kijsirikul, N. Cercone, & T.-B. Ho (Eds.), Advances in Knowledge Discovery and Data Mining (pp. 572–579). Springer. https://doi.org/10.1007/978-3-642-01307-2_55

Taud, H., & Mas, J. (2018). Multilayer perceptron (MLP). Geomatic approaches for modeling land change scenarios, 451-455.

Teufel, S. (1999). Argumentative zoning information extraction from scientific text. Ph.d.Thesis University of Edinburgh.

Teufel, S., Siddharthan, A., & Batchelor, C. (2009). Towards Domain-Independent Argumentative Zoning: Evidence from Chemistry and Computational Linguistics. Proceedings of the 2009 Conference on Empirical Methods in Natural Language Processing, 1493–1502. https://aclanthology.org/D09-1155

Teufel, S., Siddharthan, A., & Tidhar, D. (2006). Automatic Classification of Citation Function. Proceedings of the 2006 Conference on Empirical Methods in Natural Language Processing, 103–110.

Toulmin, S. (2003). The uses of argument (Updated ed). Cambridge University Press.

Tzoumpas, K., Estrada, A., Miraglio, P., & Zambelli, P. (2024). A data filling methodology for time series based on CNN and (Bi) LSTM neural networks. IEEE Access, 12, 31443-31460.

Usmani, S., & Shamsi, J. A. (2023). LSTM based stock prediction using weighted and categorized financial news. PloS one, 18(3), e0282234.

Walton, D. N., Reed, C., & Macagno, F. (2008). Argumentation schemes. Cambridge University Press.

Wang, H.-C., Cheng, J.-W., & Yang, C.-T. (2022). SentCite: A sentence-level citation recommender based on the salient similarity among multiple segments. Scientometrics. https://doi.org/10.1007/s11192-022-04339-0

Wang, J., Zhu, L., Dai, T., & Wang, Y. (2020). Deep memory network with Bi-LSTM for personalized context-aware citation recommendation. Neurocomputing, 410, 103–113. https://doi.org/10.1016/j.neucom.2020.05.047

Yang, A., & Li, S. (2018). SciDTB: Discourse Dependency TreeBank for Scientific Abstracts. CoRR, abs/1806.03653. http://arxiv.org/abs/1806.03653

Yang, L., Zhang, Z., Cai, X., & Dai, T. (2019). Attention-Based Personalized Encoder-Decoder Model for Local Citation Recommendation. Computational Intelligence and Neuroscience, 2019, 1232581. https://doi.org/10.1155/2019/1232581

Yang, L., Zhang, Z., Cai, X., & Guo, L. (2019). Citation Recommendation as Edge Prediction in Heterogeneous Bibliographic Network: A Network Representation Approach. IEEE Access, 7, 23232–23239. https://doi.org/10.1109/ACCESS.2019.2899907

Yin, M. J., Wang, B., & Ling, C. (2024). A fast local citation recommendation algorithm scalable to multi-topics. Expert Systems with Applications, 238, 122031.

Yu, B., Wang, J., Guo, L., & Li, Y. (2020). Measuring Correlation-to-Causation Exaggeration in Press Releases. Proceedings of the 28th International Conference on Computational Linguistics, 4860–4872. https://doi.org/10.18653/v1/2020.coling-main.427

Zhang, C., Liu, L., & Wang, Y. (2021). Characterizing references from different disciplines: A perspective of citation content analysis. Journal of Informetrics, 15(2), 101134. https://doi.org/10.1016/j.joi.2021.101134

Zhang, Y., & Yang, Q. (2021). A survey on multi-task learning. IEEE Transactions on Knowledge and Data Engineering, 34(12), 5586-5609.

Zhang, J., & Zhu, L. (2022). Citation recommendation using semantic representation of cited papers' relations and content. Expert Systems with Applications, 187, 115826. https://doi.org/10.1016/j.eswa.2021.115826